# Quantifying the Expectation–Realisation Gap for Agentic AI Systems

## Authors

- **Sebastian Lobentanzer** ✉
  https://orcid.org/0000-0003-3399-6695 · slobentanzer
  Institute of Computational Biology, Computational Health Center, Helmholtz Center, Munich, Germany; German Center for Diabetes Research, Munich, Germany

✉ — Correspondence possible via GitHub Issues or email to Sebastian Lobentanzer <sebastian.lobentanzer@helmholtz-munich.de>.

# Abstract

Agentic AI systems are deployed with expectations of substantial productivity gains, yet rigorous empirical evidence reveals systematic discrepancies between pre-deployment expectations and post-deployment outcomes. We review controlled trials and independent validations across software engineering, clinical documentation, and clinical decision support to quantify this expectation–realisation gap. In software development, experienced developers expected a 24% speedup from AI tools but were slowed by 19%—a 43 percentage-point calibration error. In clinical documentation, vendor claims of multi-minute time savings contrast with measured reductions of less than one minute per note, and one widely deployed tool showed no statistically significant effect. In clinical decision support, externally validated performance falls substantially below developer-reported metrics. These shortfalls are driven by workflow integration friction, verification burden, measurement construct mismatches, and systematic variation in who benefits and who does not. The evidence motivates structured planning frameworks that require explicit, quantified benefit expectations with human oversight costs factored in.

# Introduction

Agentic AI systems—autonomous software agents that plan, reason, and execute multi-step tasks with limited human oversight—are being adopted across software engineering, clinical medicine, and customer operations with the expectation of transformative productivity gains. Vendor announcements routinely promise multi-minute time savings per encounter, double-digit percentage speedups, or near-expert-level decision accuracy. Procurement and investment decisions follow these expectations, committing substantial resources before deployment-grade evidence is available.

Yet, as we explore below, a growing body of controlled trials and independent external validations reveals that realised outcomes frequently fall short of pre-deployment expectations. We term this discrepancy the *expectation–realisation gap*. It reflects systematic patterns in how agentic systems interact with human workflows, how performance is measured, and how benefits are distributed across user populations.

Understanding and quantifying this gap is a prerequisite for responsible deployment. Without structured, quantified expectations that account for real-world integration costs, organisations risk over-investing in systems that deliver marginal gains at best or impose net costs at worst. This review synthesises the strongest available empirical evidence for the expectation–realisation gap across three domains—software engineering, clinical documentation, and clinical decision support— identifies the mechanistic drivers of shortfalls, and argues that structured planning frameworks with explicit benefit quantification are necessary to close the gap between aspiration and reality.

# Evidence from controlled trials

## Software engineering copilots

The sharpest illustration of the expectation–realisation gap comes from a randomised controlled trial conducted by METR (Model Evaluation & Threat Research) on 16 experienced open-source developers working in their own mature repositories across 246 real tasks [1]. Before each task, participants forecast that AI assistance would reduce their completion time by 24%. The measured outcome was a 19% *increase* in completion time—a 43 percentage-point calibration error on the time-change scale, and a complete reversal in direction. Tasks took approximately 56% longer than developers expected. Intriguingly, AI-assisted developers also estimated 20% reduction in completion time *after* they had performed the task; the opposite of what had happened. Economics experts (N=34) and machine learning experts (N=54) overestimated the expected speedup even more dramatically, with 39% and 38%, respectively.

This result contrasts instructively with a controlled trial of GitHub Copilot on developers recruited via Upwork, where [treated participants](#) completed a standardised, self-contained programming task 56% faster (95% CI 21–89%) [2]. In that experiment, participants' self-estimated productivity gains averaged approximately 35%, meaning they *underestimated* the realised speedup. Of note, the task here was the same for all participants; the development of an HTTP server in JavaScript. The divergence between these two trials can be explained by task complexity. On constrained, well-defined tasks, AI coding tools can exceed expectations, particularly for less experienced professionals; the study notes that the benefit was greater for less experienced developers [2]. In contrast, in high-context, real-world repositories, the same class of tools can impose net costs that even senior developers fail to anticipate. These findings are the first instance of [heterogeneous treatment effects](#) of agentic AI (see the [section on heterogeneity](#)).

Field experiments at Microsoft and Accenture provide complementary information: developers assigned to Copilot completed 12.9–21.8% more pull requests per week at Microsoft and 7.5–8.7% more at Accenture, though the authors emphasise imprecision and threats to inference including low compliance and organisational confounds [3]. However, these throughput metrics do not account for code quality: independent security analyses find that 32.8% of Python and 24.5% of JavaScript snippets generated by Copilot are flagged with security issues [4], and Copilot can replicate known-vulnerable code patterns at rates around 33%, although it improves on unassisted human developers [5]. Productivity gains that increase time for human oversight (such as review, remediation, and incident risk) are not net gains.

## Clinical documentation agents

Ambient AI scribes—systems that listen to clinical encounters and generate draft documentation—represent one of the most actively deployed categories of agentic AI in healthcare. Vendor procurement narratives frequently frame benefits in terms of "minutes saved per encounter"; for instance, Microsoft publicised "5 minutes saved per clinician per encounter on average" for its DAX Copilot product [6].

Evidence from controlled trials partly contradicts these claims. A randomised controlled trial at UCLA across 238 physicians in 14 specialties compared two commercial ambient scribe tools (DAX and Nabla) against usual care, with approximately 24 000 encounters per arm [7]. Nabla reduced time-in-note by 9.5% relative to control (95% CI −17.2 to −1.8; P=0.02), while DAX showed no statistically significant effect (−1.7%, 95% CI −9.4 to +5.9; P=0.66). Partial adoption was a key contextual factor: the tools were used in only approximately 30–34% of visits, and roughly 15% of treatment-group physicians never used their assigned scribe at all. Clinicians did not strongly endorse that generated notes were "at least as good as my own"; and "occasional" clinically significant inaccuracies potentially led to a significant loss of time due to oversight activities [7].

A peer-matched cohort study of DAX in an integrated delivery system (99 providers, 12 specialties) found documentation EHR time fell from 5.3 to 4.5 minutes per patient—a saving of approximately 46 seconds—while after-hours EHR time *worsened* significantly, suggesting time-shifting rather than uniform savings [8]. A pre/post study of the Abridge ambient listening tool across 332 physicians confirmed sub-minute savings: mean time in notes per note fell from 5.1 to 4.2 minutes (difference 57 seconds, 95% CI 29–85) [9]. In this study, adoption increased from 15% to 50% of physicians in the study span of 8 weeks; however, the total number of notes created by the scribe only increased from 5% to 15% at the end of the study period.

The perception–reality mismatch that was found in software engineering studies was replicated in a study of 252 physicians: 86.5% *perceived* that their documentation time had decreased, yet there was no overall association between perceived reductions and objectively measured time changes (OR 0.975, P=0.144) [10]. The objective effect was modest: each 10 percentage-point increase in AI scribe usage was associated with approximately 30 seconds lower documentation time per scheduled hour (P<0.001).

## Clinical decision support

Where the preceding sections documented gaps in *time-savings* expectations, clinical decision support reveals equally stark expectation–realisation gaps in *accuracy and quality*—the benefit dimensions most directly linked to patient outcomes. The Epic Sepsis Model, widely implemented across US hospitals, was externally validated in a large academic health system (38 455 hospitalisations) with an area under the receiver operating characteristic curve (AUROC) of 0.63 (95% CI 0.62–0.64), while Epic previously reported AUROC values of 0.76–0.83 [11]. At an operational alert threshold, the model achieved only 33% sensitivity (missing two thirds of septic patients), raising questions about clinical utility at scale.

In oncology decision support, IBM publicised concordance rates as high as 96% for Watson for Oncology in lung cancer cases relative to a multidisciplinary tumour board [12]. A subsequent peer-reviewed retrospective study in Korea found strict concordance of 48.9% for colon cancer, with "acceptable" concordance of 65.8% and strong heterogeneity by patient age (concordance dropping to approximately 20% among patients aged 70 and older) [13]. This discrepancy reflected both definition dependence—concordance rose substantially when "for consideration" was treated as concordant—and local constraint mismatches in guidelines, reimbursement, and patient demographics that prevent cross-site transferability. In both cases, the expectation–realisation gap follows the same pattern as for time savings: internally reported metrics set expectations that external, deployment-grade evaluation cannot reproduce.

## Why expectations overshoot

The empirical evidence points to three recurrent mechanistic drivers that explain why expectations systematically exceed realised outcomes, consistent with planning fallacies widely reported in the psychology literature [14,15].

**Workflow integration friction and partial adoption.** Agentic AI systems do not operate in isolation; they must integrate into existing workflows, tools, and team practices. Clinical scribe evaluations repeatedly show partial adoption—tools used in a minority of encounters, with non-trivial drop-off over time [7,8]. Even when per-use effects are real, intention-to-treat estimates are attenuated by low compliance, and the practical benefit to an organisation depends on the adoption rate actually achieved, not the rate assumed during procurement. This is not always a simple, temporary onboarding issue; the UCLA RCT's 30–34% utilisation rate was observed over the full study period [7].

**Verification and review burden.** Agentic systems generate outputs that require human verification, and this verification cost is rarely accounted for in pre-deployment projections. In the METR software engineering trial, the net slowdown occurred because the time spent reviewing, debugging, and integrating AI-generated code exceeded the time saved in initial generation [1]. In clinical documentation, neutral ratings on note quality and "occasional" clinically significant inaccuracies indicate non-trivial editing and review work that partially or fully offsets time-in-note reductions [7]. The DAX cohort's simultaneous reduction in documentation time and *increase* in after-hours EHR time concretely shows a shift of effort, as opposed to alleviation [8].

**Measurement construct mismatch.** Pre-deployment expectations are often framed in metrics that do not correspond to what deployment-grade evaluations actually measure. Vendor claims of "minutes saved per encounter" refer to broader workflow impacts, while trial outcomes measure "time-in-note"—one slice of documentation burden [7]. Developer-reported model performance (AUROC 0.76–0.83 for Epic's sepsis model) reflects evaluation choices that can systematically inflate apparent performance relative to development goals [11]. The gap between lab-task performance and field performance in software copilots is a measurement construct problem at its core: bounded tasks estimate *tool capability under low-context load*, while field trials estimate *net productivity under realistic verification and integration costs* [1,2].

These three drivers interact and compound over time, fueling inflated expectations. Workflow integration depends on existing competence: experienced professionals can restructure their work around agentic tools while less experienced users lack the mental models to do so effectively. Verification burden scales inversely with expertise: a senior developer can spot a flawed code suggestion quickly, whereas a junior developer may accept it uncritically or spend disproportionate time reviewing it. And measurement construct mismatches extend to the time horizon of measurement itself: short-term productivity metrics cannot capture costs that materialise only over longer periods. For instance, the level of knowledge in the workforce is a slowly developing phenomenon relative to the speed at which agentic technologies are introduced. A randomised controlled trial in higher education found that students who used ChatGPT as a study aid scored significantly lower on a surprise retention test 45 days later (57.5% vs 68.5%; Cohen's d = 0.68) [16], suggesting that cognitive offloading can trade immediate task completion for degraded durable learning. A parallel study in software engineering found that AI use impaired the users' conceptual understanding, code reading, and debugging abilities, without delivering significant efficiency gains on average [17]. Ignoring these interactions when planning the deployment of an agentic system will systematically overestimate its net benefits.

## Heterogeneity as the default

Across every domain reviewed, treatment effects are not uniform. They are systematically moderated by baseline user efficiency, task complexity, and local context. This implies that development of systems useful in practice requires careful planning that respects treatment heterogeneity.

In clinical documentation, objective time savings from AI scribes concentrate among physicians with higher baseline documentation inefficiency; efficient documenters derive minimal benefit [10]. In customer support, a field study of 5,172 agents found an average 15% productivity increase from a generative AI assistant, but gains were heavily concentrated among less experienced and lower-skilled workers, while the most experienced agents saw smaller gains and occasional quality declines [18]. This heterogeneity is not only observed inter-individually but also at the intra-individual level; given a single agent, gains from AI adoption are larger for relatively rare tasks, where human users have less baseline training and experience [18]. In software engineering, the METR trial specifically selected experienced developers working in familiar repositories—precisely the population most likely to have optimised their workflows already—and this is the population that was slowed [1].

**In summary, there is currently no stable, globally positive treatment effect for agentic AI.** Average headline figures (whether from vendors, lab trials, or even well-designed field studies) will systematically misrepresent the benefit realised by any specific user, team, or organisation. Planning that relies on average expected gains without modelling who benefits and who does not will over-invest in low-yield deployments and under-invest in targeted high-yield ones.

## Implications for structured planning

The evidence reviewed here converges on a clear conclusion: pre-deployment expectations for agentic AI systems are poorly calibrated, and the resulting *expectation–realisation gap* is large enough to undermine investment decisions, deployment strategies, and trust. This is not an argument against agentic AI—the evidence also shows that real gains exist in specific contexts and for specific user populations. It is an argument for *structured planning that takes the gap seriously*.

Several design principles follow directly from the empirical patterns. First, **benefit expectations must be explicit and quantified across all relevant dimensions**, rather than framed as vague promises of efficiency. The contrast between "5 minutes saved per encounter" marketing and sub-minute measured reductions illustrates what happens when time-based expectations lack precision [7]; the gap between developer-reported AUROC (0.76–0.83) and externally validated AUROC (0.63) for the Epic Sepsis Model shows the same pattern for accuracy metrics [11]. Second, **expectations should capture dual perspectives**—what users expect to gain and what developers assess as technically feasible. Miscalibration occurs on both sides: developers overshoot in internal validation; users overshoot in self-forecasts. The estimate–reality mismatch in the METR study, which persisted even after implementation, illustrates the cognitive biases at play [1]. Third, **human oversight costs must be deducted** from projected benefits. Every controlled trial reviewed here shows that verification, review, and cleanup absorb a substantial fraction of the gross time savings; ignoring this yields unrealistic net benefit estimates. Fourth, **outcome metrics must link back to initial expectations** in the same units and at the same level of granularity, enabling direct comparison rather than post hoc rationalisation. Ideally, plans are formalised early, versioned, and archived, in order to facilitate these later comparisons. Fifth, **heterogeneity should be modelled explicitly** by specifying which user populations and task types are expected to benefit, rather than assuming uniform effects.

One way to operationalise these principles is the [Agentic Automation Canvas (AAC)](#), a structured framework for designing, governing, and documenting agentic automation projects [19]. The canvas captures user expectations as quantified benefit metrics across five dimensions—time, quality, risk, enablement, and cost—with baseline values, confidence levels from both user and developer perspectives, and explicit accounting for human oversight. This multi-dimensional structure reflects the evidence reviewed here: the expectation–realisation gap manifests not only in time savings (as in software engineering and clinical documentation) but equally in accuracy and quality metrics (as in clinical decision support), and planning frameworks must accommodate all of these. The canvas formalises the bidirectional contract between stakeholders that the evidence reviewed here shows is necessary: without structured mechanisms for surfacing and testing expectations, the gap between aspiration and reality will persist.

## Conclusion

The expectation–realisation gap in agentic AI systems is empirically documented, directionally consistent, and mechanistically explainable. Across software engineering, clinical documentation, and clinical decision support, pre-deployment expectations systematically overestimate realised benefits in deployment settings, regardless of whether they hail from user forecasts, vendor claims, or developer-reported metrics. The drivers are complex, but not mysterious: workflow integration friction, verification burden, measurement construct mismatches, and treatment effect heterogeneity are observable, predictable, and addressable in principle.

Closing the expectation–realisation gap requires moving from *ad hoc* expectation-setting to structured, quantified planning that accounts for real-world integration costs, models heterogeneity across user populations, and links outcome measurement directly to initial benefit projections. It also requires a mature interdisciplinary approach; mismatch from psychological bias cannot be countered by computer science methodology, and building better AI models will not solve all socio-technical problems in deployment and adoption. The alternative to closing the gap is to continue relying on artificial benchmark results, marketing claims, and intuitive forecasts. In all likelihood, this will perpetuate a cycle of over-promise and under-delivery that erodes trust in systems that, when properly targeted and governed, can deliver genuine value.

# Glossary of experimental terms

This review draws on evidence from controlled experiments across multiple domains. The following terms, standard in medicine, economics, and the social sciences, are used throughout.

## Randomised controlled trial (RCT)

An experimental design in which participants are randomly assigned to either a *treatment* group or a *control* group. Random assignment ensures that observed differences in outcomes can be attributed to the intervention rather than to pre-existing differences between groups. In this review, RCTs include trials of AI coding assistants, ambient clinical scribes, and clinical decision-support models.

## Treatment and intervention

The *treatment* (or *intervention*) is the condition being evaluated in an experiment—for example, giving developers access to an AI coding assistant or equipping physicians with an ambient scribe. Participants who receive the treatment are referred to as *treated participants* or the *treatment group*.

## Control

The *control* condition is the comparison group that does not receive the intervention. Control participants continue with their usual workflow, providing a baseline against which the treatment's effect is measured.

## Treatment effect

The *treatment effect* is the measured difference in outcomes between the treatment group and the control group. For example, if treated developers complete tasks 20% faster than control developers, the treatment effect is a 20% reduction in completion time. A treatment effect can be positive (the intervention helps), negative (it hurts), or null (no detectable difference).

## Intention-to-treat analysis

*Intention-to-treat* (ITT) analysis includes all participants as originally assigned—whether or not they actually used the tool. This preserves the validity of randomisation and reflects real-world conditions, where not everyone who is offered a tool adopts it. ITT estimates are typically smaller than per-use estimates because non-adopters dilute the measured effect.

## Heterogeneous treatment effects

*Heterogeneous treatment effects* means that the size (or direction) of the treatment effect varies across subgroups. For instance, less experienced developers may benefit substantially from AI assistance while expert developers see no gain or a net slowdown. Recognising heterogeneity is critical for deployment planning: an average treatment effect can mask the fact that some users benefit greatly while others are harmed.